\pdfoutput=1
\documentclass[aps,prl,twocolumn,superscriptaddress,floatfix]{revtex4-2}

\usepackage[T1]{fontenc}
\usepackage{graphicx}
\usepackage[caption=false]{subfig}

\usepackage[normalem]{ulem}
\usepackage{dcolumn}
\usepackage{bm}

\usepackage{MnSymbol}
\usepackage{physics}
\usepackage{xcolor}

\usepackage[bottom]{footmisc}
\usepackage{amsfonts}
\usepackage{wasysym}
\usepackage{ifsym}
\usepackage[export]{adjustbox}
\usepackage{dsfont}
\usepackage{cleveref}
\usepackage{lipsum}
\usepackage{stmaryrd}
\usepackage{MorrisIn,lettrine,Romantik}

\usepackage{cancel}
\usepackage{booktabs}
\usepackage{array}
\usepackage{appendix}

\newcommand{\cE}{\mathcal{E}}
\newcommand{\cB}{\mathcal{B}}

\newcommand{\kap}{\kappa}

\let\eth\relax
\newcommand{\eth}{\text{\dh}}
\newcommand{\ethb}{\bar{\text{\dh}}}

\begin{document}

\definecolor{yaleblue}{rgb}{0.06, 0.3, 0.57}
\newcommand\BOnote[1]{{\color{yaleblue}{\bf BO: #1}}}

\newcommand{\ijm}[1]{{\color{magenta}\bf ijm: #1}}

\newcommand{\jpm}[1]{{\color{olive}\bf jpm: #1}}

\newcommand{\bs}[1]{{\color{teal}\bf BS: #1}}


\title{Gravitational Memory Correlators and BMS Symmetry}

\author{Enrico Herrmann}
\affiliation{Mani L. Bhaumik Institute for Theoretical Physics,
	University of California at Los Angeles,\\
	Los Angeles, CA 90095, USA}

\author{Ian Moult}
\affiliation{Department of Physics,
    Yale University,
    New Haven, CT 06511, USA}

\author{Brett Oertel}
\affiliation{Department of Physics,
    Yale University,
    New Haven, CT 06511, USA}

\author{Julio Parra-Martinez}
\affiliation{Institut des Hautes Etudes Scientifiques, 91440 Bures-sur-Yvette, France}

\author{Biswajit Sahoo}
\affiliation{School of Physical Sciences, Indian Association for the Cultivation of Science (IACS),
2A and 2B Raja S.C. Mullick Road, Kolkata 700 032, India
}

\begin{abstract}
We compute the n-point memory and mixed energy-memory correlators for scalar decay into gravitons using the in-in formalism at leading order in perturbative quantum gravity. We obtain compact results for arbitrary n and explicitly verify the supertranslation flux balance relation between memory, energy, and Bondi mass detectors, realizing asymptotic symmetry constraints on infrared-finite observables.
\end{abstract}

\maketitle

\section{\label{sec:intro}Introduction}

\lettrine[lines=3]{G}{ravity} famously enlarges the asymptotic symmetries
of asymptotically flat spacetimes from the Poincar\'e group to the
infinite-dimensional Bondi-Metzner-Sachs (BMS) group
\cite{Bondi:1962px,Sachs:1962wk}.
These symmetries have long shaped our understanding of gravitational
radiation \cite{Ashtekar:1981bq,Ashtekar:1981hw,Ashtekar:2014zsa}.
More recently, their connection to soft-graviton theorems
\cite{Strominger:2013jfa,He:2014laa} revealed a striking link
\cite{Strominger:2014pwa} between scattering symmetries and an observable
imprint of radiation: gravitational memory
\cite{Zeldovich:1974gvh,Braginsky:1987kwo,Christodoulou:1991cr,Thorne:1992sdb}.
This perspective has inspired new memory observables
\cite{Pasterski:2015tva} and enhanced tree-level symmetry algebras
\cite{Strominger:2021mtt}.

In parallel, the observation of gravitational waves by LIGO and Virgo \cite{LIGOScientific:2016aoc,LIGOScientific:2017vwq} has spurred tremendous progress in the calculation of gravitational scattering observables \cite{Bern:2021yeh,Bern:2021dqo,Dlapa:2022lmu,Driesse:2024xad,Driesse:2024feo,Dlapa:2025biy,Bern:2025wyd,Driesse:2026qiz,Herrmann:2021lqe,Herrmann:2021tct, Cristofoli:2021vyo,Brandhuber:2023hhy,Herderschee:2023fxh,Georgoudis:2023lgf,Elkhidir:2023dco,Bohnenblust:2023qmy,Brunello:2025AnalyticWaveform,DeAngelis:2026PMWaveforms,Kavanagh:2026SFWaveforms}, including nonlinear gravitational memory \cite{Christodoulou:1991cr,Blanchet:1992br,Georgoudis:2025vkk,Oertel:2026oqv}. Related developments include logarithmic soft theorems \cite{Sahoo:2018lxl,Saha:2019tub,Sahoo:2020ryf,Alessio:2024onn,Choi:2024ygx,Karan:2026jzs} and departures from the smooth asymptotic framework \cite{Penrose:1965am,ChruscielMacCallumSingleton1995,Geiller:2022vto,Geiller:2024ryw,DeAngelis:2025vlf, Fuentealba:2024lll,Compere:2026vvi}. Furthermore,  theoretical studies of gravitational memory \cite{Flanagan:2014kfa, Flanagan:2015pxa, Nichols:2017rqr, Nichols:2018qac, Compere:2016hzt, Elkhidir:2024izo, Elkhidir:2026ahd, Bini:2024rsy, Bieri:2013ada, Garfinkle:2017fre} have been complemented by its incorporation into numerical simulations \cite{Mitman:2020pbt,Grant:2023jhd,Mitman:2024uss,Mitman:2022kwt,Mitman:2020bjf,Mitman:2021xkq}.  Finally, prospects for measuring gravitational memory are discussed in \cite{Boersma:2020gxx,Hubner:2019sly,Lasky:2016knh,Grant:2022bla,Gasparotto:2023fcg,Goncharov:2023woe}.

\begin{figure}
\centering
\includegraphics[width=0.60\linewidth]{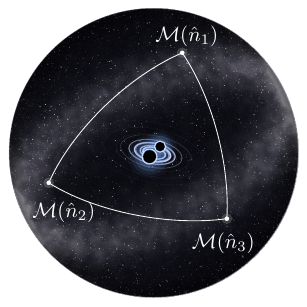}
\caption{\label{fig:intro} An illustration of a three-point memory correlator in the night sky. }
\end{figure}

It is of great theoretical interest to understand if and how asymptotic symmetries extend to a theory of quantum gravity in asymptotically flat Minkowski space. This has been extensively explored in the context of scattering amplitudes and the celestial holography program \cite{Strominger:2017zoo,He:2014laa,Kapec:2014opa,Cachazo:2014fwa,Campiglia:2014yka,Barnich:2009se,Barnich:2011mi,Kapec:2016jld,Pasterski:2016qvg,Pasterski:2017kqt,Guevara:2021abz,Raclariu:2021zjz,Pasterski:2021rjz,Pasterski:2021raf,Donnay:2022aba,Donnay:2023mrd}. A component of this question is understanding the space of observables in a potential theory of four-dimensional flat-space quantum gravity, and how these symmetries act on them. This is non-trivial for scattering amplitudes in four dimensions due to infrared divergences. In the context of quantum field theory, a particular class of asymptotic observables are in-in expectation values of detector operators, which measure asymptotic fluxes, such as energy \cite{Hofman:2008ar}. These are often referred to as energy correlators and have attracted significant recent interest \cite{Moult:2025nhu}.
Multi-point energy correlators have been computed in perturbative quantum gravity \cite{Gonzo:2020xza,Herrmann:2024yai,Chicherin:2025keq,Ruan:2026xyd}, where they provide arguably the simplest example of an infrared finite observable. Correlators of a fixed classical radiation profile factorize \cite{Gonzo:2020xza}, whereas averaging over classical radiation profiles can generate connected correlations \cite{Moynihan:2026Inclusive}. The action of the Lorentz group on these observables is well understood, and relates energy correlator configurations with different angles. However, the action of the BMS group is much more interesting, relating energy correlators to multi-point correlators of memory \cite{Moult:2025njc}. Calculating these would thus enable direct tests of the symmetries of finite observables in perturbative quantum gravity. Results beyond energy flux are limited to one-point event shapes \cite{Prabhu:talk, Himwich:2025ekg,Himwich:2026exq, Strominger:2026yrh}, and no multi-point correlator involving memory has been computed.

In this \emph{Letter}, we compute multi-point correlators of memory in perturbative quantum gravity, finding a compact solution for the $n$-point memory correlator for all $n$. Unlike energy correlators, the transcendental weight of the result does not grow with $n$.  Furthermore, we formulate the supertranslation Ward identity as a flux balance relation between memory, energy, and Bondi mass detectors and verify it using our analytic results. Our result provides a bridge between the study of asymptotic symmetries and detector operators.

\section{Memory detectors in gravity}
We consider linearized Einstein gravity with metric $\mathfrak{g}_{\mu\nu} = \eta_{\mu\nu} + \kappa\, \mathfrak{h}_{\mu\nu}$, where $\kappa = \sqrt{32 \pi G_N}$. We will construct our operators from the asymptotic graviton field
\begin{align}
\label{asymptoticgravitonfield}
h_{\mu\nu}(u,\hat n) \equiv \lim_{L\to\infty} L\mathfrak{h}_{\mu\nu}(x^\rho+Ln^\rho)\,.
\end{align}
We work in four dimensions with mostly-plus signature. The direction $\hat n$ determines the future-directed null vector $n^\mu=(1,\hat n)$, and $u=-x\cdot n$ is the retarded time.

The energy detector for the graviton can be expressed in terms of the asymptotic graviton field as
\begin{align}
\mathcal{E}(\hat n) \equiv \int_{-\infty}^\infty du : (\partial_u h_{\mu\nu}(u,\hat n)) (\partial_u h^{\mu\nu}(u,\hat n)) :\, .
\end{align}
Here we will only consider dynamical gravitons at asymptotic infinity but it is not difficult to generalise to include matter contributions. Energy detectors were constructed in gravity in \cite{Gonzo:2020xza,Herrmann:2024yai}, and their correlation functions have been computed in perturbation theory in \cite{Herrmann:2024yai,Chicherin:2025keq,Ruan:2026xyd}. The algebra of detectors of the symmetries of the Lorentz group was studied in {\cite{Gonzo:2020xza}}, and found to form an algebra {\cite{Cordova:2018ygx}}. The energy detector in gravity is a ``passive detector'', namely an eigendetector of asymptotic graviton states,  $\mathcal{E}(\hat n) |X\rangle= \sum_i k_i^0  \delta^{(2)}(\hat n ,\hat k_i) |X\rangle $, where $|X\rangle$ is an outgoing multi-graviton state.

\begin{figure}
\centering
\includegraphics[width=0.45\linewidth]{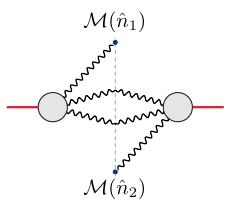}
\hfill
\includegraphics[width=0.45\linewidth]{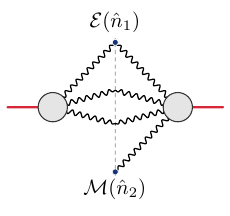}
\caption{\label{fig:mem_E_cut} Memory-Memory and Energy-Memory correlators are computed as in-in correlation functions for a scalar decaying to gravitons. The shaded blobs denote amplitudes and their conjugates.}
\end{figure}

The memory detector is the zero-frequency limit of a light-ray operator linear in the asymptotic field \cite{Strominger:2014pwa}:
\begin{equation}
\label{eq:mem_def}
\mathcal M_r(\hat n)\equiv 
\epsilon^r_{\mu\nu}(\hat n)
\lim_{\omega_0\to0^+}
\int_{-\infty}^{+\infty}du\,
\cos(\omega_0u)\,\partial_u h^{\mu\nu}(u,\hat n).
\end{equation}
Frequency-resolved light-ray operators have been studied in \cite{Korchemsky:2021htm,Korchemsky:2021okt}.
Here $r=\pm2$ is the helicity; below, $+$ and $-$ abbreviate $+2$ and $-2$. The action of the memory detector on perturbative states can be computed by simplifying Eq.~\eqref{eq:mem_def} using the stationary phase approximation:
\begin{equation}
\label{memdetstationaryphase}
\begin{aligned}
    \mathcal{M}_{r}(\hat n) =& \lim_{\omega_0\rightarrow 0^+}\frac{-\omega_0}{8\pi}(\alpha_{+r}(\omega_0 n)+\alpha^{\dagger}_{-r}(\omega_0 n)) \,.
\end{aligned}
\end{equation}
The memory detector creates or annihilates soft gravitons, so its correlators involve interferences between amplitudes with different graviton multiplicities on either side of the cut (Fig.~\ref{fig:mem_E_cut}) \cite{Kosower:2018adc,Caron-Huot:2023vxl}. Below, we compute these correlators using the Weinberg soft theorem.

\section{$n$-point memory correlators}

\begin{figure*}[tp]
 \centering
 \subfloat[\label{fig:Iplusplus}]{%
  \includegraphics[width=0.48\textwidth]{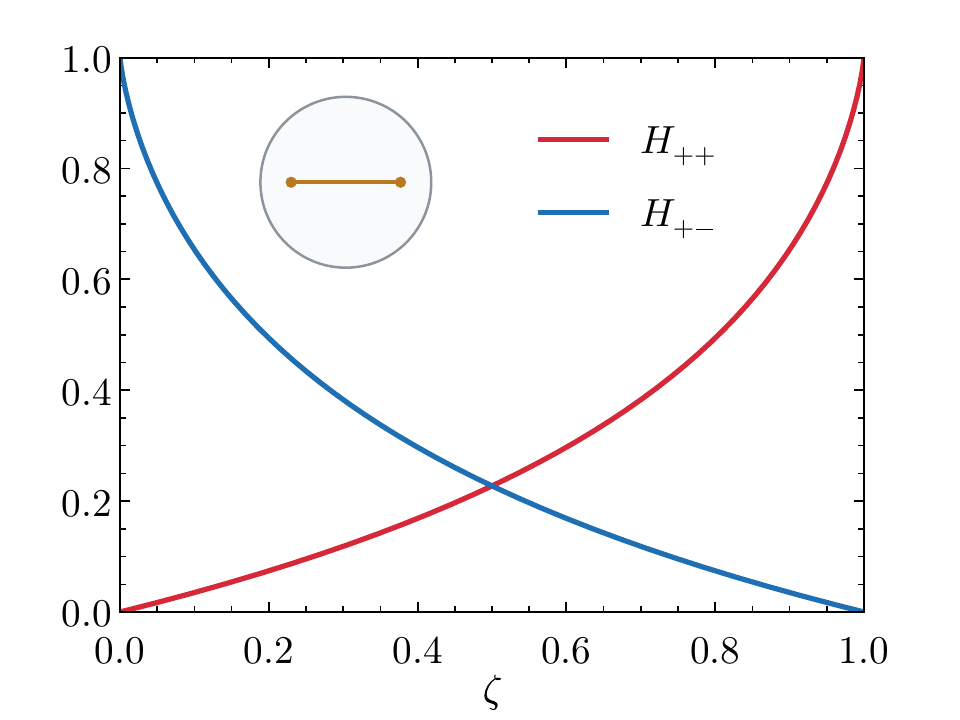}}
 \hfill
 \subfloat[\label{fig:Ithree}]{%
  \includegraphics[width=0.48\textwidth]{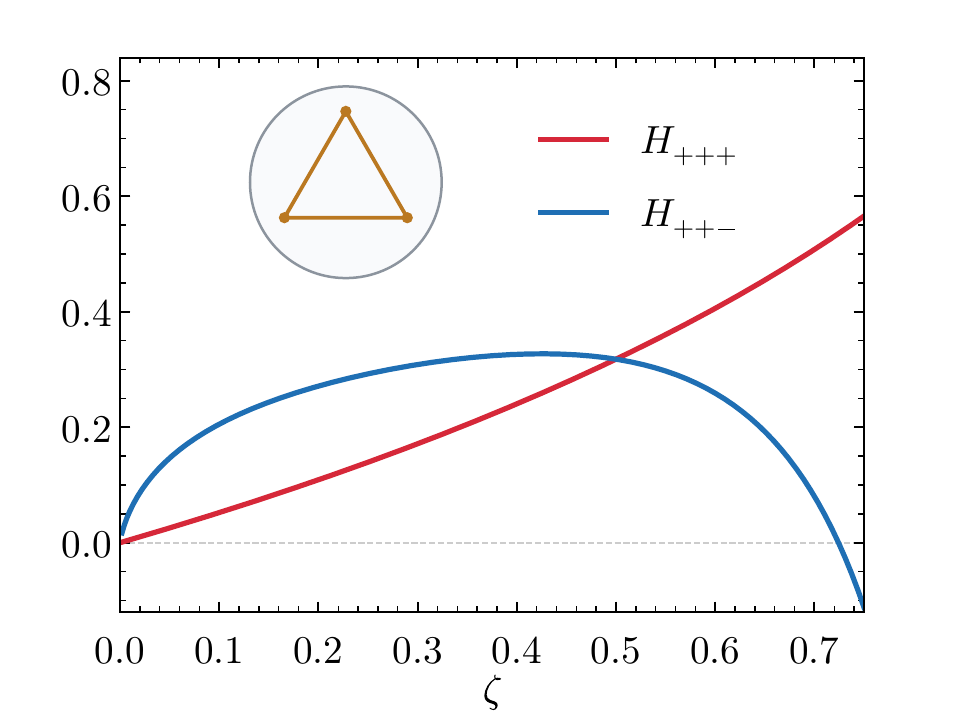}}
 \caption{(a) The helicity-stripped two-point memory functions $H_{++}$ and $H_{+-}$ as a function of $\zeta=(1-\cos\theta_{12})/2$. (b) The helicity-stripped three-point functions $H_{+++}$ and $H_{++-}$ for equal pairwise separation $\zeta$. The overall decay factor $W(\kappa M/16\pi)^n$ is divided out.}
 \label{fig:memory-two-three-point}
\end{figure*}

We consider a scalar $\chi$ of mass $M$ and momentum $P=(M,\mathbf0)$ decaying into gravitons. This gives a spherically symmetric decay and avoids forward-scattering divergences. Since minimal coupling does not allow for this, we include the interaction
\begin{align}
    \mathcal{L}_{\chi C^2}&=\frac{\chi}{\Lambda}c_{E}C_{\mu\nu\rho\sigma}C^{\mu\nu\rho\sigma} \,,
\end{align}
where $\Lambda$ has mass dimension 1, $c_{E}$ is a dimensionless EFT coefficient, and $C_{\mu\nu\rho\sigma}$ is the Weyl tensor. At leading order, the integrated two-graviton decay weight is
\begin{equation}
 W\equiv\frac{1}{2!}\sum_{\lambda_1,\lambda_2}
 \int\dd\Phi_2\,
 \left|\mathcal A^{\rm tree}_{\chi\to h_{\lambda_1}h_{\lambda_2}}\right|^2
 =\frac{\kappa^4M^8c_E^2}{32\pi\Lambda^2}.
 \label{eq:decay-weight}
\end{equation}
Here $\mathcal A^{\rm tree}$ is the decay amplitude with the momentum-conserving delta function stripped off, and $\dd\Phi_2=(2\pi)^4\delta^{(4)}(P-k_1-k_2)\prod_{a=1}^2\dd^3\mathbf k_a/[(2\pi)^3 2k_a^0]$. The factor $1/2!$ accounts for the identical final-state gravitons.
The memory operators commute when sent to null infinity using the limit in \eqref{asymptoticgravitonfield}. Each memory insertion adds a soft graviton to either the amplitude or its conjugate. At leading order, the Weinberg soft theorem \cite{Weinberg:1965nx} reduces these interference terms to the squared hard decay amplitude multiplied by soft factors. Summing over the placements of the insertions and integrating over the two-graviton phase space gives the correlator below.\footnote{This leading-soft reduction also admits a dressed-state interpretation \cite{Oertel:2026oqv}.}
We express the result in spinors. For a stereographic coordinate $z_i$ on the celestial sphere, define
\begin{equation}
  |i\rangle=\begin{pmatrix}1\\z_i\end{pmatrix},
  \quad
  |i]=\begin{pmatrix}1\\\bar z_i\end{pmatrix}\,,\quad
    |\bar i\rangle=\begin{pmatrix}-\bar z_i\\1\end{pmatrix},
  \quad
  |\bar i]=\begin{pmatrix}-z_i\\1\end{pmatrix},
  \label{eq:memory-spinor-representatives}
\end{equation}
where the barred spinors denote spinors of the antipodal point. The square and angle products are defined as
\begin{equation}
  \langle ab\rangle=a_2b_1-a_1b_2,
  \qquad
  [ab]=\tilde a_1 \tilde b_2-\tilde a_2\tilde b_1\,.
  \label{eq:memory-spinor-contractions}
\end{equation}
In terms of spinors, the angular invariant is
\begin{equation}
 \zeta_{ij}\equiv\frac{1-\cos\theta_{ij}}2
  =-\frac{\langle ij\rangle[ij]}{[i\bar i] [j\bar j]}
  =\frac{|z_i-z_j|^2}{(1+|z_i|^2)(1+|z_j|^2)}.
  \label{eq:memory-spinor-zeta}
\end{equation}

We use $\langle\cdots\rangle$ for the decay contribution to the in-in expectation value. The $n$-point memory correlator in the rest frame of the scalar is
\begin{equation}
\begin{aligned}
 &\left\langle
 \prod_{i=1}^n\mathcal M_{r_i}(\hat n_i)
 \right\rangle\\
 &\quad=W
 \left(\frac{\kappa M}{16\pi}\right)^n
 \int\frac{\dd\Omega_{\hat n_0}}{4\pi}
 \prod_{i=1}^n
 \left(
 \frac{\langle0i\rangle[0\bar i]}{[i0][i\bar0]}
 \right)^{-r_i/2}.
\end{aligned}
\label{eq:memory-angular-integral}
\end{equation}
Here $\hat n_0$ is the direction of one outgoing graviton; the other is antipodal. It can be evaluated using elementary applications of the residue theorem after decomposing into partial fractions.

\begin{figure*}[!t]
 \centering
 \subfloat[All-plus helicities ($+++$).\label{fig:memory-collinear-all-plus}]{%
  \includegraphics[width=0.48\textwidth]{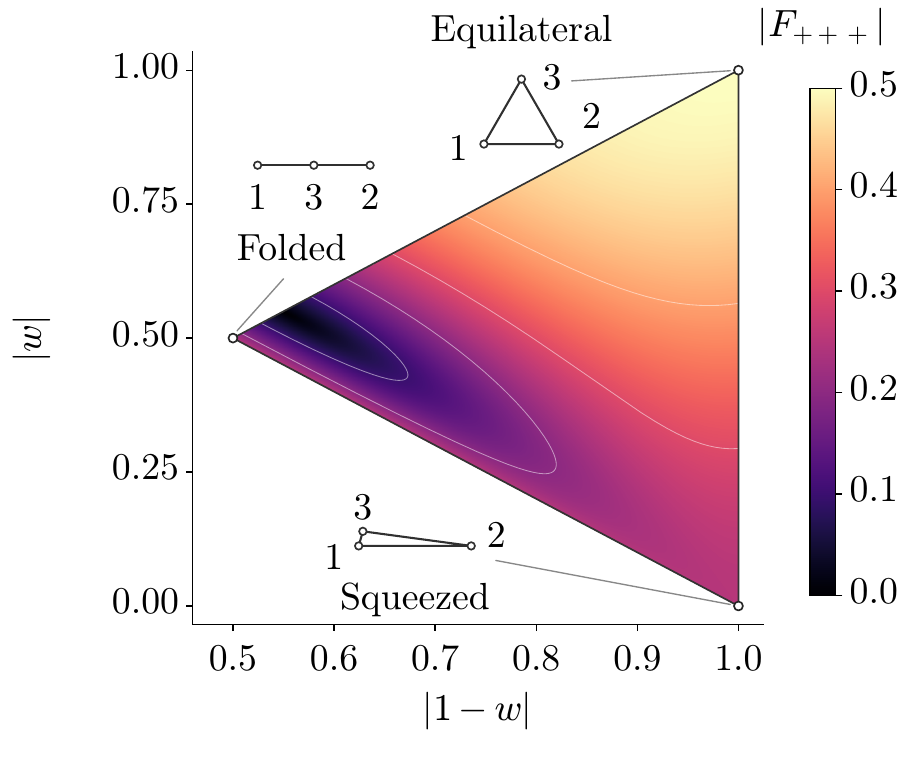}}
 \hfill
 \subfloat[Mixed helicities ($++-$).\label{fig:memory-collinear-mixed}]{%
  \includegraphics[width=0.48\textwidth]{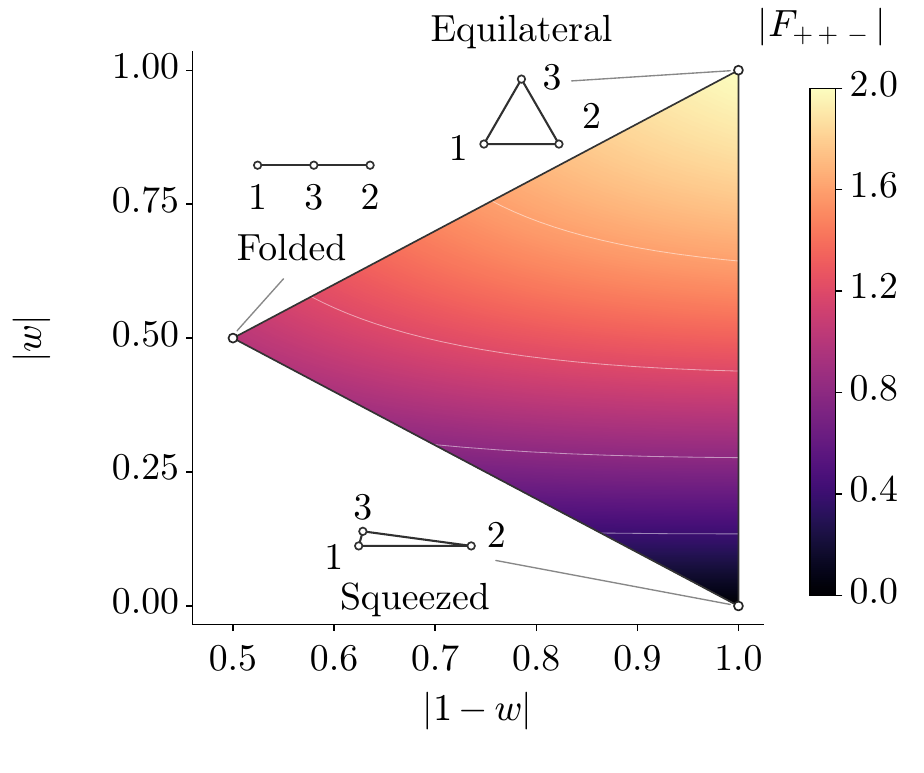}}
 \caption{Magnitudes of the leading triple-collinear shape coefficients [Eq.~\eqref{eq:memory-triple-collinear-shapes}] in the ordered domain $|w|\leq|1-w|\leq1$, $|w|+|1-w|\geq1$. Detector 3 has negative helicity in (b).}
 \label{fig:memory-collinear-ratios}
\end{figure*}

For $n\geq2$, define the dimensionless angular kernel by
\begin{equation}
 \left\langle
 \prod_{i=1}^n\mathcal M_{r_i}(\hat n_i)
 \right\rangle
 =W \left(\frac{\kappa M}{16\pi}\right)^n K_{r_1\cdots r_n}.
 \label{eq:memory-kernel-normalization}
\end{equation}
The all-plus kernel takes the compact form
\begin{equation}
\begin{aligned}
  K_{+\cdots+}
  ={}&(-1)^n\sum_{i<j}
   \left(\frac{[ij]}{\langle ij\rangle}\right)^2
   \prod_{k\neq i,j}\frac{[k\bar i][k\bar j]}
                            {\langle ik\rangle\langle jk\rangle}\\
   &\times\left[G(1-\zeta_{ij})-
   \frac1n \frac{\langle ij\rangle}{[ij]}\sum_{\ell\neq i,j}
   \frac{[i\ell][j\ell]}{[\ell\bar i][\ell\bar j]}\right]\,,
\end{aligned}
  \label{eq:memory-spinor-all-plus}
\end{equation}
where the maximal weight-1 transcendental function $G(x)$ is defined as
\begin{equation}
  G(x)=1+\frac{2x}{1-x}+\frac{2x\log x}{(1-x)^2}\,.
  \label{eq:memory-spinor-G}
\end{equation}
Every other helicity assignment is obtained from \eqref{eq:memory-spinor-all-plus} by applying the following substitution to each leg whose helicity is flipped: $|i\rangle\longmapsto|\bar i\rangle\,, ~ |i]\longmapsto|\bar i]$, and $ |\bar i\rangle\longmapsto-|i\rangle\,, ~ |\bar i]\longmapsto-|i].$ 

This is the first result for an $n$-detector correlator. It takes a structurally different form than for energy correlators. Multi-point energy correlators have been computed in gauge theory and gravity
{\cite{Chen:2019bpb,Chicherin:2024ifn,Yan:2022cye,Yang:2022tgm,Yang:2024gcn,He:2024hbb}}, and it is observed that they have transcendentality $n-1$. Since the memory correlator can be computed in the eikonal limit, the linearized structure of the propagators greatly simplifies the integrals. Its transcendental weight is one for all $n$. The result is completely finite for all angles. For $n=2$, we define the helicity-stripped functions
\begin{equation}
\begin{aligned}
 H_{++}&\equiv\left(\frac{\langle12\rangle}{[12]}\right)^2K_{++}
 =G(1-\zeta_{12}),\\
 H_{+-}&\equiv\left(\frac{\langle1\bar2\rangle}{[1\bar2]}\right)^2K_{+-}
 =G(\zeta_{12}).
\end{aligned}
 \label{eq:draft-two-stripped}
\end{equation}
Figure~\ref{fig:Iplusplus} shows these functions: $H_{++}$ vanishes in the collinear limit, while $H_{+-}$ approaches a finite value. For three detectors, we similarly define
\begin{equation}
 H_{+++}\equiv\frac{K_{+++}}{\mathcal T_{+++}},\qquad
 H_{++-}\equiv\frac{K_{++-}}{\mathcal T_{++-}},
 \label{eq:draft-three-stripped}
\end{equation}
with the unit-modulus helicity prefactors
\[
\begin{aligned}
 \mathcal T_{+++}&=\frac{[12][23][31]}{\langle12\rangle\langle23\rangle\langle31\rangle},\quad
 \mathcal T_{++-}&=\frac{[12][2\bar3][\bar3 1]}{\langle12\rangle\langle2\bar3\rangle\langle\bar3 1\rangle}.
\end{aligned}
\]
Figure~\ref{fig:Ithree} shows $H_{+++}$ and $H_{++-}$ on an equilateral slice.

The triple-collinear limit sends all three $\zeta_{ij}$ to zero at fixed ratios. Writing the stereographic coordinates as $z_1=0$, $z_2=\epsilon$, and $z_3=\epsilon w$, this is $\epsilon\to0$ at fixed $w$, with $\bar w=w^*$. The limiting side ratios are $|w|^2=\lim\zeta_{31}/\zeta_{12}$ and $|1-w|^2=\lim\zeta_{23}/\zeta_{12}$. Using $\sigma=(\zeta_{12}+\zeta_{23}+\zeta_{31})/3$, we define the complex shape coefficients by
\begin{align}
 H_{+++}&=\sigma F_{+++}+O(\sigma^2),\nonumber\\
 H_{++-}&=\sigma\log(1/\sigma)F_{++-}+O(\sigma).
 \label{eq:memory-triple-collinear-shapes}
\end{align}
Figure~\ref{fig:memory-collinear-ratios} shows the magnitudes of these generally complex shape coefficients. At the equilateral shape, $F_{+++}=1/2$ and $F_{++-}=2$.

The OPE of memory correlators has been extensively discussed  \cite{Himwich:2020rro,Arkani-Hamed:2020gyp,Kapec:2021eug,Choi:2026swo,Bissi:2024brf}. We find that in the small angle limit a pair of opposite-helicity detectors gives a term $\zeta_{ij} \log\zeta_{ij}\,,$ while in the two-point correlator a same-helicity pair vanishes as $\zeta_{ij}$, without a logarithm,
in line with predictions. Note, however, that since we consider a finite observable, the coefficient is finite. It would be interesting to develop the OPE structure of memory detectors. We believe our result contains a wealth of data for understanding the structure of memory correlators.

We can also compute the mixed energy-memory correlator, which in the rest frame takes the simple form
\begin{equation}
\begin{aligned}
 &\left\langle\cE(\hat n_0)
 \prod_{i=1}^n\mathcal{M}_{r_i}(\hat n_i)\right\rangle\\
 &\quad=W\frac{M}{4\pi} \left(\frac{\kap M}{16\pi}\right)^n
 \prod_{i=1}^n
 \left(\frac{\langle0i\rangle[0\bar i]}{[i0][i\bar0]}\right)^{-r_i/2}
 .
\end{aligned}
 \label{eq:sf-em-N}
\end{equation}
It obeys the sum rule
\begin{equation}
 \int\dd\Omega_{\hat n_0}\,
 \langle\cE(\hat n_0)\prod_i\mathcal{M}_{r_i}(\hat n_i)
 \rangle
 =M\langle\prod_i\mathcal{M}_{r_i}(\hat n_i)
 \rangle \ .
 \label{eq:em-sum-rule}
\end{equation}

The two-point energy correlator needed for the BMS flux balance check is
\begin{equation}
 \langle{\cE(\hat n_1)\cE(\hat n_2)}\rangle
 =W\frac{M^2}{32\pi^2}
 \left[\delta(\zeta_{12})+\delta(1-\zeta_{12})\right]\,.
 \label{eq:sf-eec}
\end{equation}

\section{BMS Flux Balance Laws}

The full BMS symmetry group is expected to place an infinite number of symmetry constraints on correlators of detector operators. While the Lorentz subgroup relates different configurations of the same detector, non-trivial BMS symmetries relate expectation values of different detector operators. This was investigated in \cite{Moult:2025njc}. Using our explicit results for the memory and energy correlators, we are able to verify these relations for supertranslations, which relate energy and displacement memory.

We now express the classical supertranslation flux balance law \cite{Ashtekar:1981bq,WaldZoupas2000Charges} in detector language. The corresponding relation between memory, radiated energy, and the change in Coulombic data follows from the asymptotic Einstein equations \cite{Frauendiener1992Memory}; see \cite{Compere:2019gft} for a modern treatment. The Bondi constraint can be written as
\begin{equation}
 \partial_u m_B
 =\frac14\left(D_z^2N^{zz}+D_{\bar z}^2N^{\bar z\bar z}\right)-\frac14N_{zz}N^{zz}\,,
 \label{eq:bal-local-constraint}
\end{equation}
where $m_B$ is the Bondi mass aspect and $N_{AB}=\partial_u C_{AB}$ is the news tensor, with $C_{AB}$ the tracefree Bondi shear. Integrating over retarded times, we obtain
\begin{equation}
\begin{aligned}
 &\frac{1}{2}D^zD^z\Delta C_{zz}
 +\frac{1}{2}D^{\bar z}D^{\bar z}\Delta C_{\bar z\bar z}\\
 &\quad=\frac12\int_{-\infty}^{+\infty}\dd u\,N_{zz}N^{zz}+2\Delta m_B,
\end{aligned}
 \label{eq:bal-chiral-shear}
\end{equation}
where $\Delta$ denotes the final-minus-initial difference. The integrated news term defines the energy flux operator
\begin{equation}
 \cE(\hat n)=\frac{2}{\kap^2}
 \int_{-\infty}^{+\infty}\dd u\,N_{zz}N^{zz}\,.
 \label{eq:bal-news-energy-detector}
\end{equation}
The shear divergences in \eqref{eq:bal-chiral-shear} are scalars, while the memory detectors have spin weight $\pm2$. They are related by the Newman--Penrose spin-raising and spin-lowering operators \cite{NewmanPenrose1966BMS,GoldbergEtAl1967SpinHarmonics}. In our spinor conventions,
\begin{equation}
 \eth_i=(|\bar i\rangle)^\alpha
 \frac{\partial}{\partial(|i\rangle)^\alpha},\qquad
 \ethb_i=(|\bar i])^{\dot\alpha}
 \frac{\partial}{\partial(|i])^{\dot\alpha}}\,.
 \label{eq:bal-eth-spinor}
\end{equation}
These operators give
\begin{equation}
 \begin{aligned}
 D^zD^z\Delta C_{zz}&=\frac{\kappa}{2}\,\ethb^2\mathcal{M}_{+2},\\
 D^{\bar z}D^{\bar z}\Delta C_{\bar z\bar z}
 &=\frac{\kappa}{2}\,\eth^2\mathcal{M}_{-2}.
 \end{aligned}
 \label{eq:bal-memory-left-hand-side}
\end{equation}

Finally, we express $\Delta m_B$ as a Bondi mass detector. We use the null pair $n^\mu=(1,\hat n)$ and $\bar n^\mu=(1,-\hat n)$, normalized by $n\cdot\bar n=-2$, and define:
\begin{equation}
 \cB(\hat n)\equiv-\frac{1}{\kap^2}
 \int_{-\infty}^{+\infty}\dd u\,
 \partial_u\left[\lim_{r\to\infty}r^3
 C_{n\bar n n\bar n}(u,r,\hat n)\right].
 \label{eq:bal-curvature-detector}
\end{equation}
The shear--news term vanishes at both endpoints, so
\eqref{eq:bal-curvature-detector} reduces to
\begin{equation}
\begin{aligned}
 \cB(\hat n)&=\frac{8}{\kap^2}\Delta m_B(\hat n)
\end{aligned}
 \label{eq:bal-aspect-detector}
\end{equation}
which measures the difference of the final and initial Coulombic data.

We are therefore able to express the BMS flux balance relation for supertranslations as the detector operator statement (note the two halves on the LHS are equal since the magnetic memory vanishes)
\begin{equation}
\ethb^2\mathcal{M}_{+2}=\eth^2\mathcal{M}_{-2}=\frac{\kappa}{2}\big(\cE+\cB\big)\,,
 \label{eq:bal-helicity-law}
\end{equation}
relating memory, energy and Bondi mass detectors. This relation should hold inside correlators with an arbitrary product of additional detector operators.

Using the explicit results for the memory, energy-memory, and energy-energy correlators, we can verify that this relation is indeed true. To do so with an additional energy detector, we require the following distributional identity to hold
\begin{equation}
 \begin{aligned}
 \ethb_i^2
 \left(\frac{[i0][i\bar0]}{\langle0i\rangle[0\bar i]}\right)
 &=\eth_i^2
 \left(\frac{\langle0i\rangle[0\bar i]}{[i0][i\bar0]}\right)\\
 &=4\pi\left[\delta^{(2)}(\hat n_i,\hat n_0)
 +\delta^{(2)}(\hat n_i,-\hat n_0)\right]-2\,,
 \end{aligned}
 \label{eq:bal-unit-response}
\end{equation}
with $\int\dd\Omega_{\hat n}\,\delta^{(2)}(\hat n,\hat n')=1$, which is crucial to reproduce contact terms in the energy correlator distribution. Applying it to the mixed memory-energy correlator gives
\begin{align} \label{eq:bal-ME-distribution}
& \ethb_1^2\langle{\mathcal{M}_{+2}(\hat n_1)\cE(\hat n_2)}\rangle
 =\eth_1^2\langle{\mathcal{M}_{-2}(\hat n_1)\cE(\hat n_2)}\rangle\nonumber\\
 &=W\frac{\kap M^2}{16\pi}
 \left[\delta^{(2)}(\hat n_1,\hat n_2)
 +\delta^{(2)}(\hat n_1,-\hat n_2)\right]
 -W\frac{\kap M^2}{32\pi^2}\,,
 \end{align}
 which is precisely the energy-energy correlator and the difference in Bondi mass. One can show that this relation holds with $n$ additional memory detectors, providing a highly non-trivial check on its structure.


The flux relation \eqref{eq:bal-helicity-law} can also be inverted. Firstly, since $\ethb^2\mathcal{M}_{+2}$ contains only $\ell\geq 2$ parts, the $\ell \leq 1$ components of $\cE+\cB$ vanish (this is just energy-momentum conservation). Next, $\ethb^2$ acts invertibly on the $\ell\geq 2$ harmonics with coefficient $\sqrt{(\ell-1)\ell(\ell+1)(\ell+2)}$ \cite{GoldbergEtAl1967SpinHarmonics}. Thus, since the memory contains no magnetic part ($\ethb^2\mathcal M_{+2}=\eth^2\mathcal M_{-2}$) the two-point memory correlator follows from \eqref{eq:sf-eec}. With $\hat n_1$ at the north pole and $\hat n_2$ at zero azimuth,
\begin{equation}
\begin{aligned}
&\langle\mathcal{M}_{+2}(\hat n_1)\mathcal{M}_{\pm2}(\hat n_2)\rangle\\
 =&W\left(\frac{\kappa M}{16\pi}\right)^2\sum_{\ell\geq2,\ \ell\ \rm even}
 \frac{4(2\ell+1)\,d^{\,\ell}_{2,\mp2}(\theta_{12})}{(\ell-1)\ell(\ell+1)(\ell+2)}\,,
 \label{eq:mem-harmonics}
\end{aligned}
\end{equation}
with $d^{\,\ell}_{m m'}$ the Wigner functions. All coefficients are positive, inherited from the energy-energy correlator, and their $\ell^{-3}$ tail is consistent with the $\zeta_{12}\log\zeta_{12}$ term of $\langle\mathcal{M}_{+}\mathcal{M}_{-}\rangle$.

\section{Conclusion}

In this \emph{Letter}, we computed multi-point displacement-memory correlators for scalar decay into gravitons at leading order in perturbative quantum gravity. We obtained a compact analytic expression for arbitrary detector number and helicities, finite at all angular separations and of transcendental weight one independently of $n$. Together with the mixed energy-memory correlators, these results explicitly verify the supertranslation flux balance relation between memory, energy, and Bondi mass detectors. They demonstrate how asymptotic symmetries constrain the angular dependence of finite observables through relations among distinct detector correlators.

Beyond leading order, the central questions are whether memory correlators remain infrared finite and whether the supertranslation flux balance relation holds in the presence of quantum corrections. The cancellation of infrared divergences demonstrated for gravitational energy correlators at one loop \cite{Chicherin:2025keq} provides a precedent.

More broadly, our results suggest studying gravitational detectors as families connected by asymptotic symmetries. Work on higher-spin flux laws and light-ray symmetry algebras \cite{Geiller:2024bgf,Himwich:2025ekg} motivates extending this perspective to higher memories, whose detectors may organize into symmetry multiplets. This extension requires determining which higher detectors exist as quantum operators and whether their definition requires renormalization. Our calculation realizes this perspective for supertranslations and provides an explicit starting point for understanding the space of gravitational observables through its asymptotic symmetries.

\begin{acknowledgments}
{\bfseries Acknowledgements} We thank Thibault Damour, Steve Giddings, Carlo Heissenberg, Mina Himwich, Murat Kologlu, David Kosower, Sabrina Pasterski, Monica Pate, Ashoke Sen, Ali Seraj and Alexander Zhiboedov for useful discussions.
I.M. and B.O. are supported by the DOE Early Career Award DE-SC0025581, the Sloan Foundation, and the Simons Collaboration on Confinement and QCD Strings. The work of J.P.-M. is supported by the European Research Council (ERC) under the European Union’s Horizon 2020 research and innovation programme, grant agreement ERC-StG-101221094 GravitaS. BS acknowledges the hospitality of IHES during an IRN:QFS supported Short Scientific Visit, where this project began.
\end{acknowledgments}




\bibliography{memory_refs_v2}

\end{document}